\documentclass{ws-ijmpe}
\usepackage{color}
\usepackage{url}
\usepackage{subcaption}
\usepackage{hyperref} 
\usepackage[super,compress]{cite}
\begin{document}

\markboth{Dexheimer et al.}{Phase Transitions in Neutron Stars}

\catchline{}{}{}{}{}
%%%%%%%%%%%%%%%%%%%%%%%%%%%%%%%%%%%%%%%%%%%%%%%%%%%%%%%%%%%%%%%%%%%%

\title{Phase Transitions in Neutron Stars}

\author{V. Dexheimer $^1$\footnote{vdexheim@kent.edu}\ , L. T. T. Soethe $^2$, J. Roark $^1$, R. O. Gomes $^3$, S. O. Kepler $^2$, S. Schramm $^3$}

\address{$^1$ Department of Physics, Kent State University, Kent OH 44242 USA \\
$\,$
$^2$ Instituto de F\'{i}sica, Universidade Federal do Rio Grande do Sul, Av. Bento Gon\c{c}alves 9500\\
Porto Alegre, Rio Grande do Sul 91501-970, Brazil \\
$\,$
$^3$  Frankfurt Institute for Advanced Studies, Frankfurt am Main, Germany}

\maketitle

%\begin{history}
%\received{Day Month Year}
%\revised{Day Month Year}
%\accepted{Day Month Year}
%\comby{(xxxxxxxxxx)}
%\end{history}

\begin{abstract}
In this paper we review the most common descriptions for the first order phase transition to deconfined quark matter in the core of neutron stars. We also present a new description of these phase transitions in the core of proto-neutron stars, in which more constraints are enforced so as to include trapped neutrinos. Finally, we calculate the emission of gravitational waves associated with deconfinement phase transitions, discuss the possibility of their detection, and how this would provide information about the equation of state of dense matter.
\end{abstract}


%\ccode{PACS numbers:26.60.Dd, 26.60.Kp, 04.30.-w, 04.40.Dg}

%\tableofcontents

\section{Introduction}

Neutron stars (NS's) are a natural laboratory for the study of dense matter. Their interiors cover a large range of densities going from about $1$ g/cm$^3$ in the atmosphere to about $10^{15}$ g/cm$^3$ 
--- a number density of about $1$ baryon per fm$^{3}$ --- in the stellar core. The latter value corresponds to a volume per baryon
less than the size of a nucleon, implying that at such densities baryons overlap. This can be understood as a strong indication of deconfined quark matter in the interior of NS's. From a stability point of view, it was long ago establish that  3-flavored quark matter could be more energetically stable than hadronic matter \cite{Bodmer:1971we,Witten:1984rs}  and, more recently, the same was shown for 2-flavored quark matter \cite{Holdom:2017gdc}\,.

After the first work proposing pure quark stars in 1970 \cite{Itoh:1970uw}\,, Glendenning started the discussion of conserved charges in hybrid hadronic-quark stars in 1992. He highlighted the fact that, if allowed, a mixture of phases will take place when first order phase transitions take place \cite{Glendenning:1992vb}\,. Consequently, the pressure is not constant in the extended mixture, as the concentrations of the substances change together with the chemical potentials associated with the constraints (two in this case, global baryon number and global electric charge). This became known as Gibbs construction, as equilibrium conditions require the Gibbs free energy per particle (i.e. the baryon chemical potential), temperature and pressure to be equal in both phases within the mixture. Finally, a volume fraction of each substance can be calculated at any point in the mixture which fulfills the globally required constraints. Note that previous works had studied mixtures of hadronic and quark phases, although not in the context of astrophysics ~\cite{Lukacs:1986hu, Heinz:1987sj, Glendenning:1992vb}\,.

If charge neutrality is imposed locally in each phase, there is no mixture of phases. The pressure is constant, in the sense that it relates to the value of the chemical potential associated with the conserved quantity (which is different in each phase). The Gibbs free energy per particle, temperature, and pressure are still equal in both phases. This is known as Maxwell construction, as it does not allow the pressure to change as a function of number density. But, as a result, constant pressure means that this region does not occupy a physical space under the influence of gravity in a star (unlike the mixture of phases). More recently, this discussion appeared again in the literature in a more general form and using the terms congruent and non-congruent referring to Maxwell and Gibbs constructions, including two  or more constraints in the context of astrophysics and heavy-ion collision physics \cite{Hempel:2009vp,Hempel:2013tfa,Roark:2018uls}\,. 

The determination of the way phase transitions take place in nature, with local or global charge neutrality (and in the latter also the extent of mixed phase), depends directly on the surface tension between the two phases. For deconfinement phase transitions, surface tension has been calculated, but shown to be model dependent \cite{Alford:2001zr,Voskresensky:2002hu,Maruyama:2007ey,Maruyama:2007ss,Palhares:2010be,Pinto:2012aq,Yasutake:2013sza,Lugones:2013ema,Garcia:2013eaa,Lugones:2016ytl}\,. The description of hybrid stars under both scenarios of global and local charge neutrality has been studied in many works in the past, and also recently used to constrain even further the equation of state (EoS) of nuclear matter in attempts to reproduce tidal deformability measurements from neutron star mergers \cite{Bandyopadhyay:2017dvi,Paschalidis:2017qmb,Annala:2017llu,Nandi:2017rhy,Most:2018hfd,Burgio:2018yix,Tews:2018iwm,Alvarez-Castillo:2018pve}\,. In this paper, we review some of the points related to phase transitions already raised in our previous works, but focusing on their relation with the possibility of identifying such phase transitions through the detection of gravitational waves (GW's).

We have recently and definitively entered the age of gravitational wave astrophysics with the discoveries of black hole and NS mergers made by the LIGO and Virgo collaborations \cite{GW1,GW1NS}\,. The next runs of the interferometers will be able to detect GW's of smaller and smaller amplitudes, raising the possibility of detecting even more subtle events \cite{Prospects2018}\,, such as the ones described in the section 4 of this work. This requires prior knowledge of the signal waveform, making the identification of possible sources and wave frequencies a relevant problem in the detection strategy. Using the EoS's presented in this paper, we estimate the initial amplitude and frequency of the gravitational waves (GW's) emitted by a NS that undergoes a phase transition going from a purely hadronic star to a hybrid one with the same number of baryons. Note that this is different from what has been recently presented in Ref.~\cite{Graeff:2018czm} or previously in Ref.~\cite{Drago:2015cea}\,, where different EoS's were used to generate purely hadronic and purely quark branches. Finally, we estimate the decay of GW amplitudes with time for some selected pulsars.

\section{Microscopic Description}

As matter in the inner core of NS's is very dense but strongly interacting, it cannot be currently described by first principle theories. Alternatively, we can rely on effective models, which after being calibrated to work in the desired regime of energies, can produce reliable results concerning the matter EoS and associated particle population. For this purpose, we choose the Chiral Mean Field (CMF) model, which is based on a non-linear realization of the SU(3) sigma model \cite{Papazoglou:1998vr}\,. It is an quantum relativistic model that describes hadrons (nucleons and hyperons) and $3$ light flavors of quarks interacting via meson exchange, as a way to describe the attractive and repulsive components of the strong force \cite{Dexheimer:2009hi,Hempel:2013tfa}\,. The model is constructed to be chirally-invariant, in a manner similar to the linear-sigma model, as the particle masses originate from (instead of being modified by) interactions with the medium and, therefore, decrease at high densities/temperatures. The non-linear realization refers to the kind of chiral transformation imposed, which has the pseudo-scalar mesons as parameters. This setup results in a framework in which there is no distinction between left- and right-handed space and, therefore, in a larger freedom in the calculation of the mesonic couplings.  The mesons included are the lowest mass ones that are scalar iso-scalar, vector iso-scalar, scalar iso-vector, and vector iso-vector (with and without hidden strangeness)~.

\begin{figure}[t!]
\centerline{\includegraphics[width=10cm]{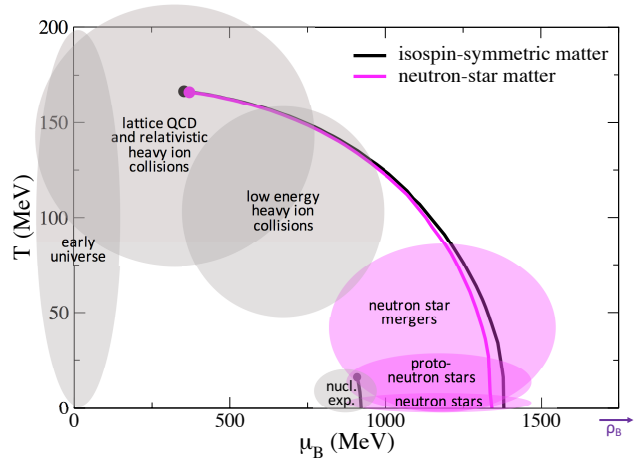}}
\caption{QCD phase diagram (temperature vs. baryon chemical potential) resulting from the CMF model with lines indicating first-order phase transition coexisting lines. The line on the bottom stands for the nuclear liquid-gas transition, while the lines on the top-right stand for the chiral-symmetry restoration/quark deconfinement transitions, The dots mark the respective critical end-points. The shaded regions exemplify relevant scenarios.}
\end{figure}

After applying the mean-field theory approximation,  the hadronic coupling constants of the model were calibrated to reproduce the vacuum masses of baryons and mesons, and were fitted to reproduce nuclear constraints for isospin symmetric matter (together with the symmetry energy) at saturation with reasonable values for the hyperon potentials. The quark coupling constants were constrained using lattice QCD data at zero baryon chemical potential \cite{Ratti:2005jh,Rossner:2007}\,, as well as information about the the remaining QCD phase diagram for isospin asymmetric and symmetric matter. The latter include the point where the coexistence line ends at the zero-temperature axis and the position of the critical point \cite{Fodor:2004nz}\,, among others. As a consequence, this formalism reproduces the nuclear liquid-gas phase transition as well as the deconfinement/chiral symmetry restoration phase transitions expected to be found in the QCD phase diagram, as shown in Fig.~1. As a final test, we have used perturbative QCD (PQCD) results, calculated by taking into account beta equilibrium and charge neutrality  \cite{Kurkela:2016was}\,, in order to determine until which density/chemical potential our model is valid. We found that our model is fully consistent with PQCD in the whole regime of densities achieved inside NS's and proto-neutron stars (PNS's) \cite{Kurkela:2016was,Roark:2018uls}\,.

The lines in Fig.~1 represent first-order transitions and the dots mark the critical end-points. Isospin-symmetric matter refers to zero-isospin matter with zero net strangeness, as the one created in heavy-ions collisions or any nuclear experiment performed in the laboratory. NS matter stands for charged neutral matter in chemical equilibrium,  such as the one inside the core of neutron stars. The shaded regions exemplify in which regimes these kinds of matter can exist. Other scenarios showed in the figure (and colored accordingly) correspond to matter created in the early universe, also isospin symmetric with zero net strangeness, and matter created in supernova explosions and neutron star mergers, also charge neutral. Chemical equilibrium is not establish immediately in supernovae and stellar mergers, but instead these events present a temporary large lepton fraction. For the case of proto-neutron star matter, a fixed lepton fraction discussion will be presented in the following.

We model PNS matter by imposing another constraint to characterize the neutrinos trapped by the dense and hot medium, lepton fraction. This is the ratio of the amount of electrons/electron neutrinos to the amount of baryons and it is fixed according to supernova simulations to be $Y_l=0.4$ \cite{Fischer:2009af,Huedepohl:2009wh}\,. The extra constraint has the effect of suppressing the hyperons (due to the presence of negatively charged electrons) and pushing the phase transition to higher chemical potentials (as it makes the quark matter EoS softer than the hadronic one), with respect to the NS case. The results from this description can be seen in the left panel of Fig.~2 for the case that all constraints (except baryon number) are enforced locally in each phase. Although this locally enforced condition might be the case for electric charge if the surface tension is large \cite{Lugones:2013ema,Garcia:2013eaa}\,, it is not the case for lepton fraction. This is because there is no long range force associated with this quantity (such as Coulomb's force for electric charge) \cite{Hempel:2009vp}\,, what leads us to refer to this case as ``forced-congruent", in which case the Gibbs free energy $\tilde{\mu}$ is not the baryon chemical potential, but a function of the lepton chemical potential $\tilde{\mu}=\mu_B + Y_l \mu_l$.

\begin{figure}[t!]
\centering
\includegraphics[width=0.495\textwidth]{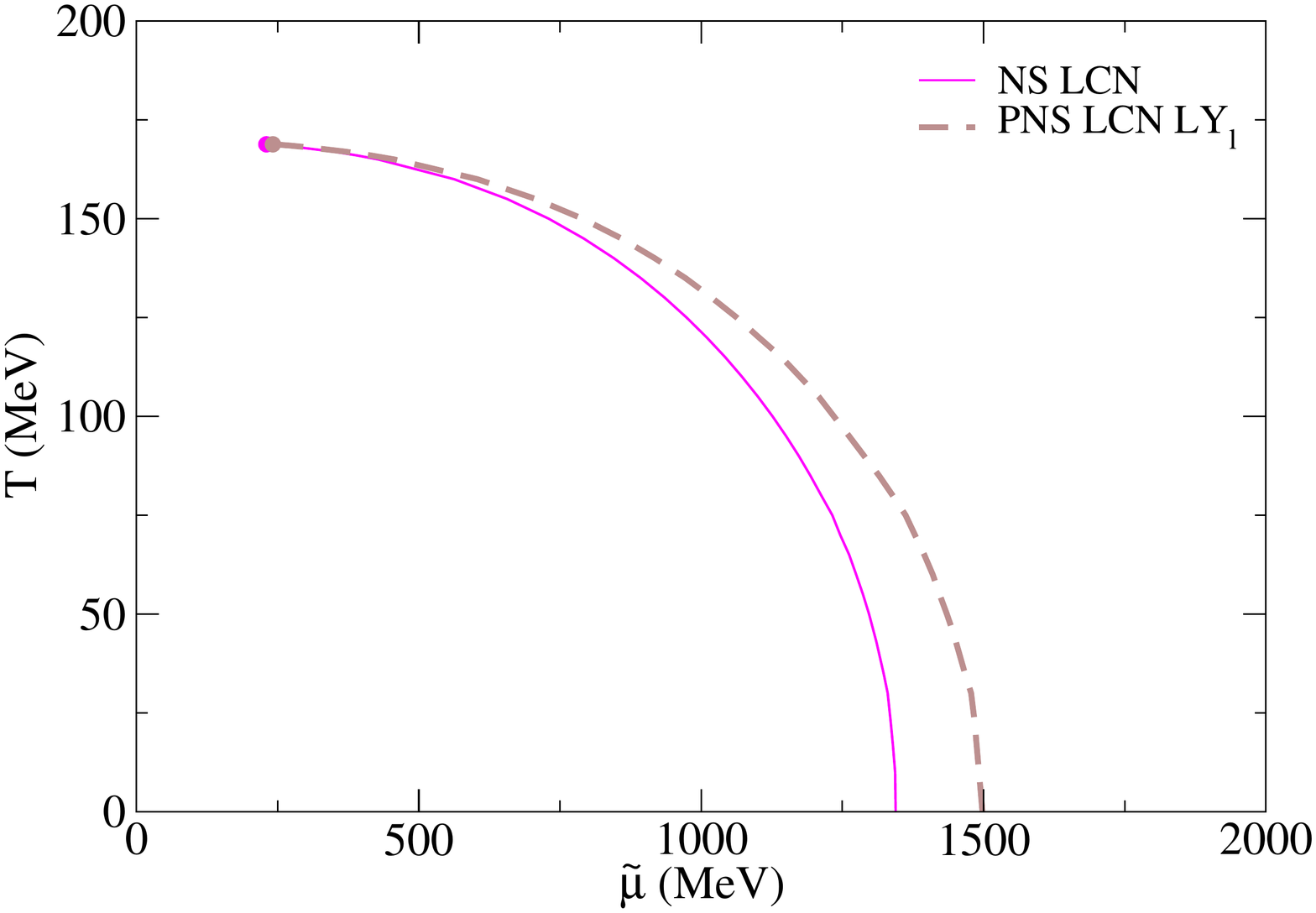} 
\includegraphics[width=0.495\textwidth]{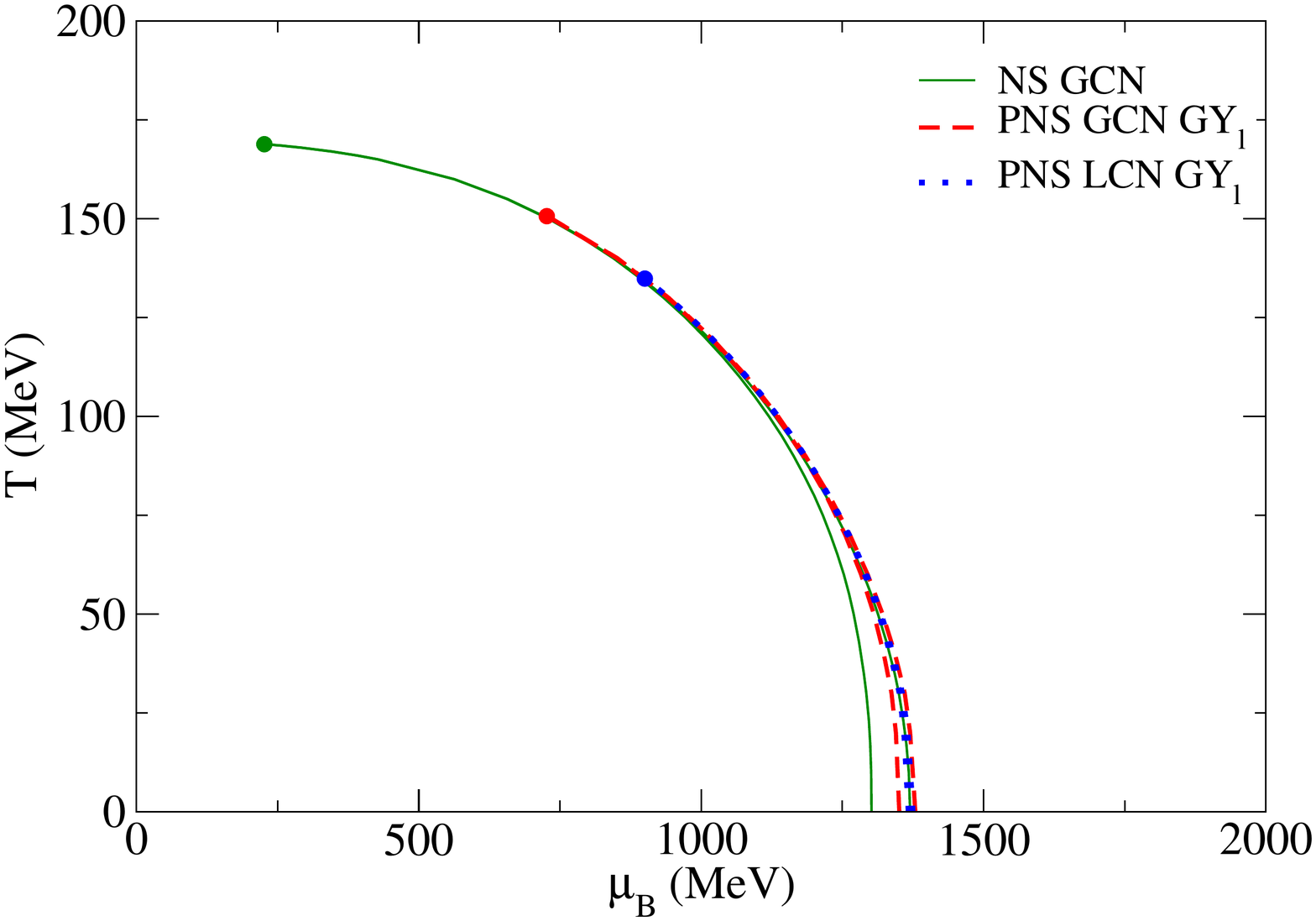} 
\caption{QCD phase diagrams, as in Fig. 1, but including one extra constraint to describe proto-neutron stars. In the left panel, electric charge neutrality and lepton fraction are enforced locally. In the right panel one or both constraints are enforced globally, creating regions with mixtures of phases.}
\end{figure}

The right panel of Fig.~2 illustrates what happens if one or two constraints (in addition to baryon number) are allowed to be conserved globally. In this description, mixtures of phases appear, although PNS matter possess much smaller mixtures of phases than those of NS matter (i.e., they extend through much smaller ranges of chemical potentials and smaller ranges of densities). In the case of global lepton fraction conservation, specially when electric charge neutrality is constrained locally, the mixtures of phases become so narrow at large temperatures that it become numerically impossible to find them. In practice, this would mean that these mixtures of phases would not impact significantly any stellar properties. Note that finite size effects tend to shrink the size of mixtures of phases even further \cite{Yasutake:2015xda,Wu:2017xaz}\,.

\section{Macroscopic Description}

Although thermal energy is negligible in NS's, this is not the case for PNS's, as they can reach tens of MeV temperature in their centers \cite{Burrows:1986me,Pons:1998mm}\,. To simulate that, we add the additional (local) constraint of fixed entropy density per baryon density $S_B=2$ in our PNS EoS's. It results in a temperature gradient in stars that, as a consequence of fixing entropy per baryon locally and not globally, has a small (practically negligible) jump across the phase transition, as discussed in detail in Ref.~\cite{ournewpaper}\,. 

\begin{figure}[t!]
\centering
\includegraphics[width=0.495\textwidth]{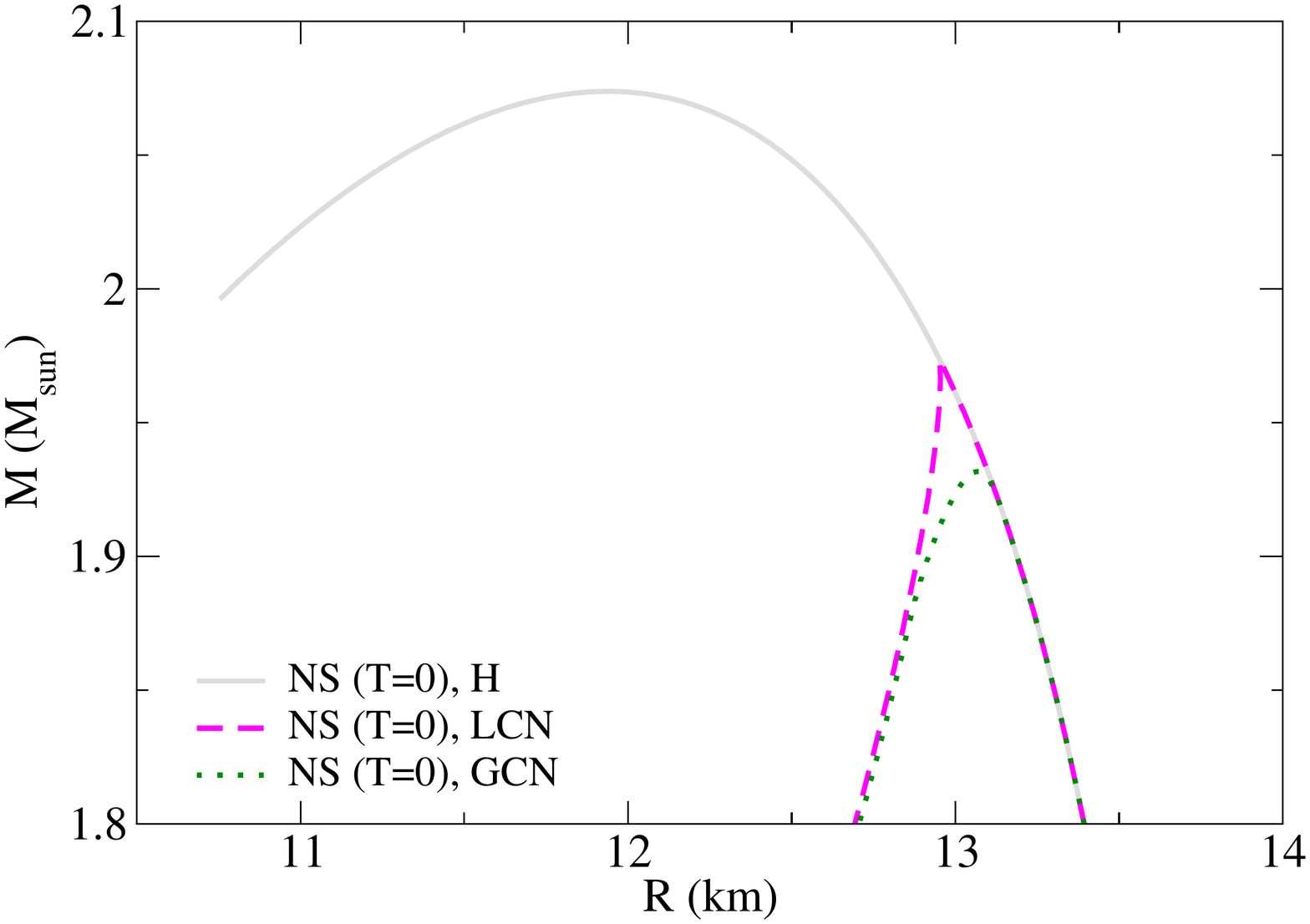} 
\includegraphics[width=0.495\textwidth]{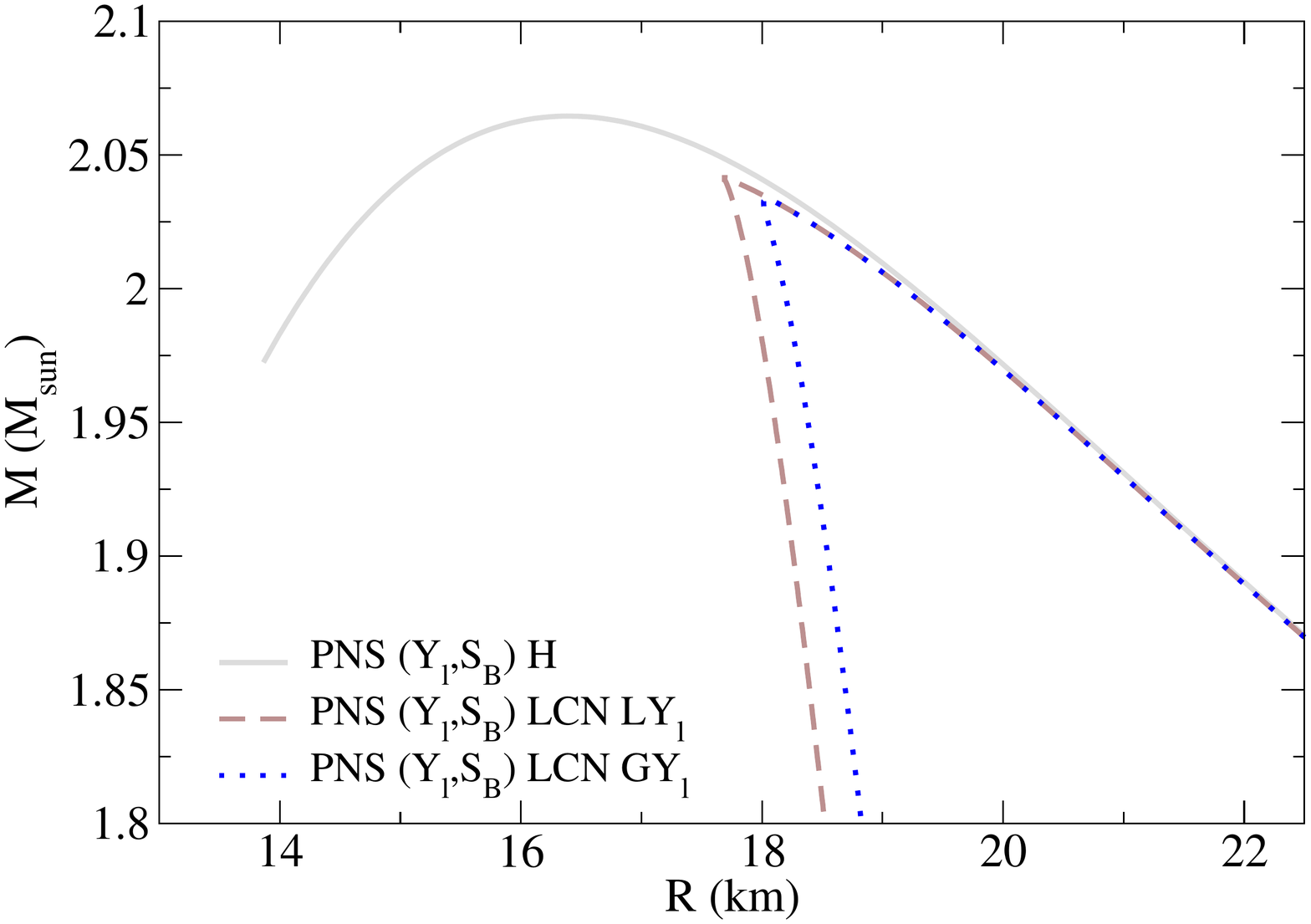} 
\caption{Solution of TOV equations showing stellar masses and radii for neutron stars (left panel) and proto-neutron stars with fixed entropy density per baryon density (right panel). The following cases are shown: when quark matter is artificially suppressed (H), when quark matter is allowed and electric charge neutrality is constrained locally (LCN) or globally (GCN) and lepton fraction is constrained locally (LCN) or globally (GCN).}
\end{figure}

Next, we use our different NS and PNS EoS's in the Tolman-Oppenheimer-Volkoff (TOV) equations to find a family of stellar solutions for each EoS, as shown in Fig.~3. For NS's, we show three curves, for hadronic matter only (H) and with a first order phase transition assuming local charge neutrality (LCN) or global charge neutrality (GCN). In the case with the local constraint, stars that reach the threshold central density for the phase transition are unstable but, in the case with the global constraint, there is an extended mixture of phases that reaches about $2$ km of radius in the most massive stable star. For PNS's, we show again three curves, for hadronic matter only (H), with a first order phase transition assuming local charge neutrality and lepton fraction (LCN LGY$_l$), and with a first order phase transition assuming local charge neutrality and global lepton fraction (LCN GY$_l$). In the case with both local constraints, stars that reach the threshold central density for the phase transition are unstable but, as before, in the case with the global constraint, there is an extended mixture of phases that reaches about $1$ km of radius in the most massive stable star. The case in which all constraints are conserved globally for PNS's is not shown in the figure, as it is very demanding numerically and does not differentiate dramatically from the LCN GY$_l$ case.

It is important to note that, in all hybrid PNS cases, there are quarks present in stable stars. This is because the CMF model allows for the existence of soluted quarks in the hadronic phase and soluted hadrons in the quark phase at finite temperature. This is discussed in detail in Ref.~\cite{ournewpaper}\,. Regardless, quarks will always give the dominant contribution in the quark phase, and hadrons in the hadronic phase and the phases can be distinguished from one another though their order parameters. We assume that this inter-penetration of quarks and hadrons (that increases with temperature) is indeed physical, and is required to achieve the crossover transition known to take place at small chemical potential values \cite{Aoki:2006we}\,.

\section{Gravitational Wave Emission due to Phase Transition}

Even though only GW's from NS's in a binary system have been detected until 2018 \cite{GW1,GW1NS}\,, it is predicted that isolated NS's could also irradiate detectable GW's through different processes. In particular, newly formed neutron stars that reach high enough central densities in their cores could undergo a phase transition to deconfined quark matter. In this case, the new hybrid configurations would be more compact (than their purely hadronic counterparts) but have the same number of baryons. Such dense metastable hadronic stars could be formed, for example, by the merging of two low mass neutron stars, two white dwarfs, or a combination of both \cite{Foucart:2015gaa,Zenati:2018gcp,Rueda:2018ban}\,. 

The conversion from a purely hadronic to a hybrid star would decrease the star's gravitational mass  $\Delta M_{G}$ and, therefore, also gravitational energy  $\Delta M_{G} c^{2}=(M_{G}^{\text{Hyb}}-M_{G}^{\text{Had}})c^{2}=\Delta E_{T}$.  Among other things, this energy can excite pulsating modes of the star. NS's have a large number of distinct vibrational modes, being the fundamental ($f$) mode, in general, the one that radiates most mechanical energy \cite{Kokkotas1997,LindblomDetweiler1983}\,. A star of mass $M$ and size $R$ has a natural GW frequency of ${f=(1/4\pi)\sqrt{3GM/R^3}}$ \ \cite{Schutz2000}\,.  So, considering typical NS values, we expect to detect waves in the range of $1\text{-}3~\text{kHz}$ when using the prescription from Ref.~\cite{Hughes2014} and considering that a deconfinement phase transition occurs. Because it is a sudden event, this phenomenon is usually classified as a burst, although the damping of the oscillation of the star's surface may in some cases last for years, as we shall discuss later.

Assuming the quadrupole moment of an arbitrary mass distribution $Q_{ij}$, the energy lost via gravitational radiation is given by the time derivative of the energy (see, e.g., Ref.~\cite{Landau1975}):
\begin{equation}
 -\frac{\text{d} {E}}{\text{d}t} = \frac{\text{G}}{45\text{c}^{2}} \left( \frac{\partial ^{3} Q_{ij}}{\partial t ^{3}} \right) ^{2} \text{,}
\end{equation}
which provides the mean luminosity of the gravitational wave emitted $L_{GW}$. We are interested in the fundamental mode, which is characterized by being a surface mode between the star interface and its surroundings \cite{Kokkotas1997}\,. Then, by conceiving a non-radial axisymmetric oscillation in a sphere of a given radius, we can express an oscillation at its surface by describing an expansion of $r(\theta)$ Ref.~\cite{Chau1967}\,. If only linear terms are considered, the expansion is reduced to its first two terms, leading up to the relation:
\begin{equation}
 L_{GW} = \frac{2 {E}_{2}}{\tau} \text{,}
\end{equation}
where ${E}_{2}$ represents the approximation in the energy
and ${\tau}$ is the damping time scale, which is expected to be relatively large for the $f$ mode,
indicating a slow damping. Then, if we consider that most of the mechanical energy is in the $f$ mode,
the gravitational strain amplitude can be written as \cite{Pacheco2011Potential}\,:
\begin{equation} \label{h_0} %
 h_{0}= \frac{4}{2 \pi f_0 r} \left( \frac{\text{G}\Delta E}{\tau \text{c}^{3}} \right) ^ {1/2} \text{,}
\end{equation}
where $r$ is the distance to the source and $h_{0}$ the amplitude measured at distance $r$.

The frequency of the fundamental mode (in kHz), is well fitted by \cite{AnderssonKokkotas1996}\,:
\begin{equation} \label{nu} %
 f_{0} \approx 0.17 + 2.30 \sqrt{ \left( \frac{10~\text{km}}{R} \right)^{3} \left( \frac{M}{1.4\text{M}_{\odot}} \right) }
\text{,}
\end{equation}
and the damping time scale by GW emission is:\cite{Chau1967}
\begin{equation} \label{tau}
 \tau = 1.8 \left( \frac{\text{M}_{\odot}}{M} \right) \left( \frac{P^{4}}{R^{2}} \right)
\text{,}
\end{equation}
where $P$ is the period of rotation of the star (in ms).

Before using the change in gravitational stellar mass due to the deconfinement phase transition $\Delta M_{G}$ as available energy as in Eq.~\eqref{h_0}, we must keep in mind that not all the energy released in the transition is converted into mechanical energy. Instead, some of it is dissipated into thermal, shear and bulk viscosity processes. Thus, we must introduce an efficiency term $\eta$ in this relation. Early works \cite{Houser1994,Schutz1996,Marranghello2002} suggest $10^{-3} \lesssim \eta \lesssim 0.5$, while more recent calculations \cite{Passamonti2013} estimate $10^{-7} \lesssim \eta \lesssim 10^{-5}$, depending on the model used. Given the large uncertainty in this parameter and the difficulty of evaluating the best value, we  scale our relation with an intermediate value  $\eta=10^{-4}$.

In our setup, the initial amplitude of the measured GW depends on mass, radius, and rotation period of the star; the distance between the star and Earth; and also of the energy released via the phase transition. Although the TOV equations only describe spherical stars, any corrections to that due to rotation and magnetic fields would deform further the star and, therefore, cause a larger emission of GW's. For each star (calculated with a given EoS and central density), rotation period and distance of the source are additional parameters that will differentiate the initial amplitude of the GW. Here, we use data from 2572 pulsars cataloged in Ref.~\cite{Manchester2005}\,. 

The previous equations also allow us to describe the wave oscillation behavior in time,
which is given by \cite{Chau1967,Marranghello2002}\,:
\begin{equation} \label{ht}
 h(t) = h_{0} e^{-(1/\tau-i 2 \pi f_0)t}
\text{.}
\end{equation}
In the case of an interferometric detector with an arm of length $L$, $h = \Delta L/L$ is measured, where $\Delta L$ is a small change in the length $L$ caused by the GW. Using this framework, we estimate the GW amplitude when metastable hadronic stars go through deconfinement phase transitions in two cases, first, assuming stars that are cold and in chemical equilibrium (previously referred to as NS's) and, second, assuming stars to be hot and with trapped neutrinos (previously referred to as PNS's).

Fig.~4 shows the results of our estimates of $h_{0}$ in comparison with the sensitivity of different GW detectors. Note that the GW amplitude $h_{0} ~ \propto ~ \Delta E^{1/2} ~, ~ 1/r ~, ~ R^{5/2} ~, ~ 1/P^{2}$.
The quantities $\Delta E$ and $R$ vary considering all the possible NS's that can undergo a deconfinement phase transition (from the hadronic to the hybrid branch with a mixed phase in the left panel in Fig.~3) at fixed baryon number. They generate very similar results. The case without a mixed phase is not considered as it does not produce stable hybrid NS's. For PNS's or, to be more specific, stars that follow more closely PNS conditions such as finite temperature and fixed lepton fraction effects, the quantities $\Delta E$ and $R$ vary considering only massive stars from the hadronic to the hybrid branch with and without a mixed phase (see right panel in Fig.~3) that can go through a deconfinement phase transition at fixed baryon number. The case without a mixed phase produces stable stars with quarks only due to the finite temperature assumed in PNS's. Less massive stars are not considered, as they generate lower energy release upon transitioning.

\begin{figure}[t!]
\centering
\includegraphics[width=10cm]{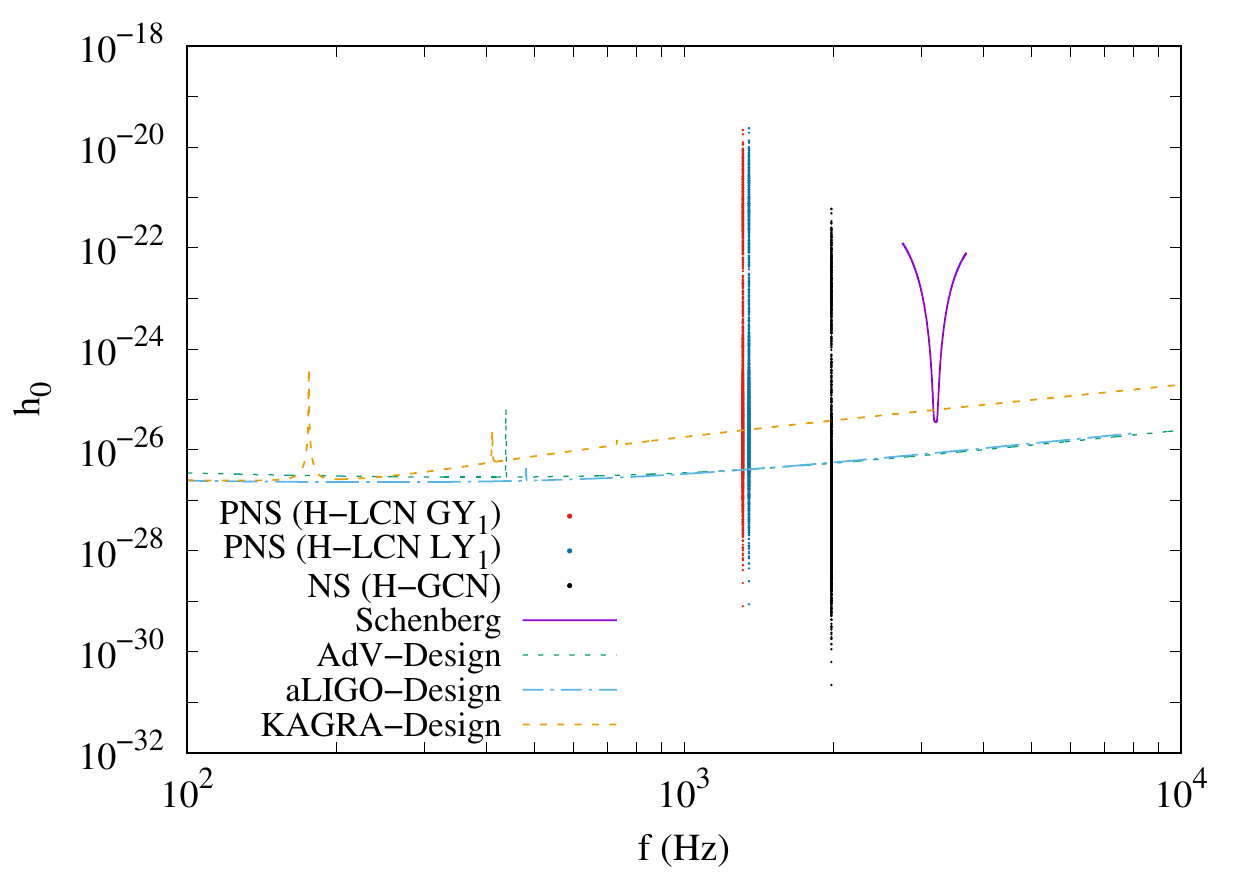} 
\caption{Initial amplitude of GW's and respective frequencies when phase transitions take place in NS's allowing the existence of a mixed phase and  in PNS's allowing for 
the existence of a mixed phase, or not. Different lines show the sensitivity of several gravitational-wave detectors \cite{Aasi:2013wya,2012JPhCS.363a2003A}\,.}
\end{figure}

\begin{figure}[t!]
\centering
\includegraphics[width=0.495\textwidth]{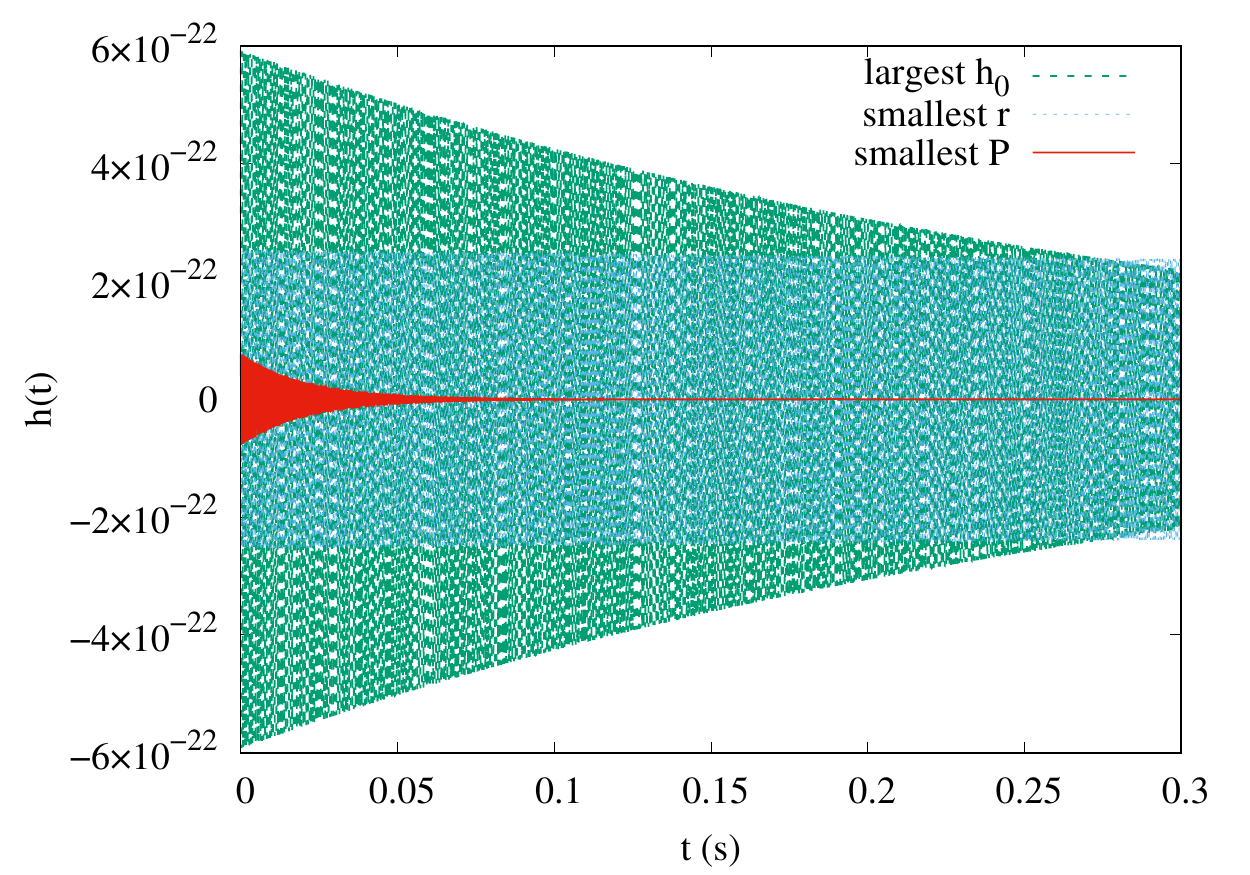} 
\includegraphics[width=0.495\textwidth]{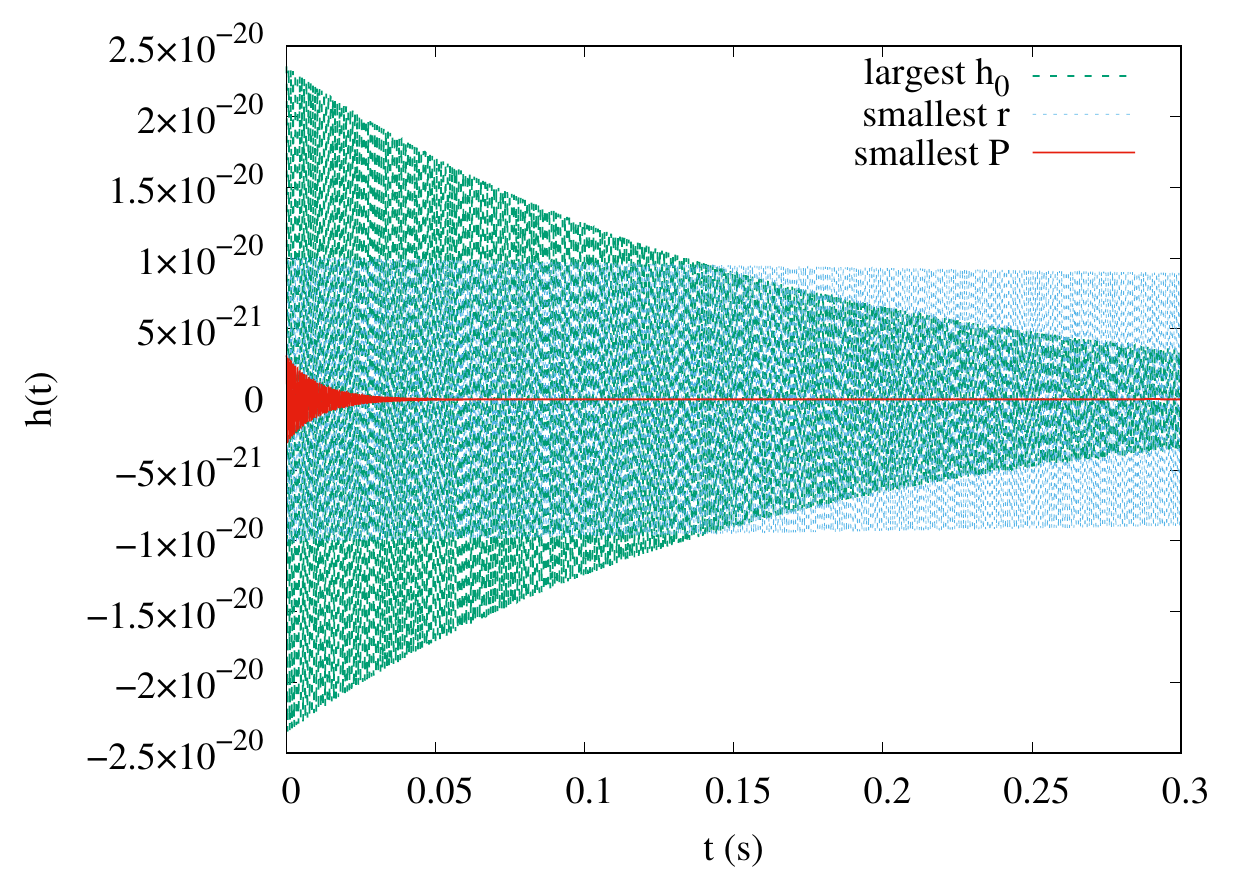} 
\caption{Decay of GW amplitude with time for three selected pulsars for NS H-LCN (left panel) and PNS H-LCN LY$_l$ (right panel) conditions, the one with the highest estimated amplitude ($h_{0}=1.9 \times 10^{-21}$), the one with the shortest rotation period ($P=1.396~\text{ms}$), and the closest to the Earth ($r=90~\text{pc}$).}
\end{figure}

As explained above, the distance $r$ and the pulsar period $P$ are varied according to available pulsar data. As a result, the pulsars with the largest $h_{0}$ are the ones with smallest periods and/or which are closest to Earth. We estimate that $h_{0}$ varies from $2.2 \times 10^{-31}$ to $5.9 \times 10^{-22}$ $(\eta/10^{-4})$ and the frequency is around $2.0$ kHz for NS's. For PNS's, $h_{0}$ varies from $8.0 \times 10^{-30}$ to $2.4 \times 10^{-20}$ $(\eta/10^{-4})$ and the frequency is around $1.3$ kHz. In the case of PNSs, the estimated initial signal for more than two thousand (83\%) pulsars are above the sensitivity limit of the LIGO and Virgo detectors, and more than one thousand (54\%) for KAGRA detector, depending on $\eta/10^{-4}$. For NS, there are six hundred (25\%) for LIGO and Virgo, and four hundred (17\%) for KAGRA. Note that the GW frequency $f ~ \propto ~ 1/R^{3/2} ~ and ~ M^{1/2}$ only, so this results could be in principle used to distinguish among different EoS's and, in the case that the stellar mass is known, the stellar radius of neutron stars. This in return could provide information about the interior of the stars.

Another point that must be highlighted is the order of magnitude of the damping time scale $\tau$, as it is proportional to the fourth power of P. Typical pulsars period values can lead to a $\tau$ of some milliseconds or several years, making the detection in the latter case very difficult, given the time of operation of the detectors. This becomes clear in Fig.~5, where we present the wave pattern for three pulsars selected from the catalog: the one with the highest estimated amplitude ($h_{0}=1.9 \times 10^{-21}$), the one with the shortest rotation period ($P=1.396~\text{ms}$) and the closest to the Earth($r=90~\text{pc}$). Here, both $r$ and $P$ are very important, but it is their combination that determines the highest amplitude of the GW signal. Pulsars with slow rotation rate, in addition to tending to decrease the amplitude of the GW, have a very long damping time and would require a very extensive operating time for the detectors. This is also the case with the pulsar with smaller distance. The left and right figure panels of Fig.~5 show results for NS H-LCN and PNS H-LCN LY$_l$ conditions, respectively. In the case of PNS's with global Y$_l$, the results are similar to what is presented in the right panel of Fig.~5, but with slightly reduced overall amplitude magnitude. Note that although pulsars with large damping time are not of interest for detection, they are of great importance for the establishment of background noise, since their signal remains practically constant for a long time.

\section{Conclusions}

In this work, we have revisited the topic of phase transitions in the interior of neutron and proto-neutron stars making use of a realistic equation of state that accounts for hadronic and quark degrees of freedom. Different possible scenarios, in which global and local charge neutrality and lepton fraction constraints were imposed. The possibility of deconfinement to quark matter in the core different stars was investigated. Although mixtures of phases extend through larger portions of cold deleptonized neutron stars, in our framework, quarks are present in hot stars even outside these mixtures. As a consequence, stars at all stages of evolution can present quarks in extended portions (if they possess large enough central density). When we compared massive but purely hadronic stars (that had the quarks suppressed artificially), and respective hybrid stars (with same number of baryons), they presented distinguished compactnesses, for all the conditions analyzed.

We then investigated the possibility of detecting GW's emitted in the case of metastable hadronic stars undergoing a deconfinement phase transition and converting to hybrid stars. This could be the case of isolated newly formed massive neutron stars formed, for example,  by the merger of low mass stars. In this case, the phase transition from a hadronic star to a more compact hybrid star with the same number of baryons would release gravitational energy and excite pulsation modes that could eventually be detected. Most of the uncertainties in our  predictions refer more to the amplitude of the detected GW's and less to their frequency, which is mainly equation of state dependent. In this way, a possible detection will be able to provide solutions to outstanding issues regarding dense matter, such as which degrees of freedom exist in the center of neutron stars, in addition to an alternative way to measure stellar radii, complimentary to electromagnetic wave measurements.

Although not unique, our scenario predicts gravitational waves that could be measured in the near future. Our results are consistent for example with the ones from Ref.~\cite{Abdikamalov:2008df}\,, which uses simple equations of state but a very sophisticated treatment of the oscillations including simulations performed using a code that solves the general relativistic hydrodynamic equations and includes rotation. We must emphasize that we used in our work an integration time of $1$ month for the GW detectors and an  efficiency of $\eta/10^{-4}$ for the relation between released gravitational energy and available energy for GW emission. Modifying $\eta$ will modify our results for the amplitudes by a factor $\eta^{1/2}$. Moreover, for simplicity we consider only the fundamental mode of oscillation, but we point out that the addition of other vibrational modes can increase the values of GW's amplitude. Ref.~\cite{Staff:2011zn}\,, for example, assesses the gravitational waveform that would result from r-mode driven spindown of magnetized neutron stars.

\section*{Acknowledgements}

Support comes from Conselho Nacional de Desenvolvimento Cient\'{i}fico e Tecnol\'{o}gico (CNPq-Brazil) and by the National Science Foundation under grant PHY-1748621.

%\begin{thebibliography}{0}

\bibliographystyle{ws-ijmpe}
\bibliography{paper}% Produces the bibliography via BibTeX.

%\end{thebibliography}

\end{document}